\documentclass[aps,pre,reprint,superscriptaddress,showpacs]{revtex4-1}
\usepackage{graphicx}
\usepackage{amsmath}
\usepackage{amssymb}

\newcommand {\e} {\varepsilon}
\def\W{\Omega}

\def\w{\omega}
\def\vp{\varphi}

\begin{document}
\title{Chimera-like states in an ensemble of globally coupled oscillators}
\author{Azamat Yeldesbay}
\affiliation{Institute for Physics and Astronomy, 
University of Potsdam, Karl-Liebknecht-Str. 24/25, 14476 Potsdam-Golm, Germany}
\author{Arkady Pikovsky}
\affiliation{Institute for Physics and Astronomy, 
University of Potsdam, Karl-Liebknecht-Str. 24/25, 14476 Potsdam-Golm, Germany}
\affiliation{Department of Control Theory, Nizhni Novgorod State University,
Gagarin Av. 23, 606950, Nizhni Novgorod, Russia}
\author{Michael Rosenblum}
\affiliation{Institute for Physics and Astronomy, 
University of Potsdam, Karl-Liebknecht-Str. 24/25, 14476 Potsdam-Golm, Germany}

\begin{abstract}
We demonstrate emergence of a complex state in a homogeneous  ensemble of globally coupled 
identical oscillators, reminiscent of chimera states in locally coupled oscillator lattices. 
In this regime some part of the ensemble forms a regularly evolving cluster, while all other units
irregularly oscillate  and remain asynchronous. 
We argue that chimera emerges because of  effective bistability which dynamically appears 
in the originally monostable system due to internal delayed feedback in individual units.  
Additionally, we present two examples of chimeras in bistable systems with frequency-dependent 
phase shift in the global coupling.
\end{abstract}
\date{\today}

\date{\today}

\maketitle

In spite of over forty years of research pioneered by 
A.~Winfree \cite{Winfree-67,*Winfree-80}
and Y.~Kuramoto \cite{Kuramoto-75,*Kuramoto-84}, 
the dynamics of globally coupled oscillator populations remains a 
challenging issue, with applications ranging from laser and Josephson junction 
arrays to problems of bridge engineering and modeling of brain 
waves~\cite{PhysRevLett.82.1963,*PhysRevE.61.2513,%
*McRobie-03,*Strogatz_et_al-05,%
*Golomb-Hansel-Mato-01,*Breakspear-Heitmann-Daffertshofer-10}. 
In addition to the well-studied self-synchronization transition, 
of particular recent interest are complex states 
between synchrony and asynchrony
\cite{vanVreeswijk-96,*Mohanty-Politi-06,*Rosenblum-Pikovsky-07}.
On the other hand, a lot of attention have attracted regimes of 
coexistence of coherence and incoherence in oscillators lattices
\cite{Kuramoto-Battogtokh-02}. 
These states, also known as ``chimeras'', have been 
addressed in numerous theoretical 
studies \cite{Abrams-Strogatz-04,*PhysRevLett.100.044105,*PhysRevLett.100.144102,%
*Bordyugov-Pikovsky-Rosenblum-10,*Laing20091569,*PhysRevLett.104.044101,*Wolfrum-11,%
*Laing_et_al-12,*Zhu_et_al-12,*PhysRevLett.110.094102}
and demonstrated in an experiment \cite{PhysRevLett.110.244102}.
Furthermore, it has been shown that already
two interacting populations of
globally coupled identical oscillators can for some initial conditions 
exhibit symmetry breaking of synchrony, so that one population 
synchronizes whereas the other remains asynchronous 
\cite{Abrams-Mirollo-Strogatz-Wiley-08,*Pikovsky-Rosenblum-08}; 
existence of such chimeras has been also 
confirmed experimentally \cite{Tinsley2012,*Martens2013}.
A natural question, addressed in this Letter, is under which 
conditions can such a 
symmetry-breaking into synchronous and asynchronous groups 
be observed in a completely homogeneous 
 globally coupled population of identical oscillators.  

In case of global coupling all oscillators are subject to 
the same force. 
Therefore, if the units are identical, one may 
expect that they should evolve similarly.
This expectation is rather natural and is indeed
true for simple systems like the standard Kuramoto 
model as well as for many other
examples from the literature. 
However, in a system of identical globally coupled chaotic maps,
K.~Kaneko observed one large synchronized cluster and a cloud of scattered units 
(see Fig 2b in~\cite{Kaneko-90b}) -- a state reminiscent of a chimera.
For periodic units such a state has been reported by  
Schmidt~\textit{et al.}~\cite{Schmidt_etal-13},
who studied nonlinearly coupled 
Stuart-Landau oscillators, see also~\cite{Sethia-Sen-13}.
These observations of identical
nonlinear elements behaving differently in spite of being driven by the same force, 
indicate presence of bi- or multistability. 
Here we demonstrate that chimera-like states naturally appear for a minimal 
generalization 
of the popular Kuramoto-Sakaguchi phase model to the case of globally coupled 
identical phase oscillators 
with internal delayed feedback, and discuss the underlying mechanism of 
dynamically sustained bistability. 

Globally coupled self-sustained oscillators can be quite generally treated
in the phase approximation~\cite{Kuramoto-84}. In the simplest case of 
identical sine-coupled units 
such an ensemble of $N$ units is described by the Kuramoto-Sakaguchi 
model~\cite{Sakaguchi-Kuramoto-86}:
\begin{equation*}
\dot\vp_k=\w +\frac{\e}{N}\sum_{j=1}^N \sin(\vp_j-\vp_k+\beta)=\w+\e\text{Im}(e^{i\beta}Ze^{-i\vp_k})\;,
\label{eq:phks}
\end{equation*}
where $\vp$ are the oscillators' phases, $\e>0$ is the coupling strength,
$\beta$ is the phase shift in the coupling, and 
$Z=Re^{i\Theta}=N^{-1}\sum_{k=1}^N e^{i\vp_k}$ is the complex Kuramoto order 
parameter (complex mean field).
The system is known to tend to 
the fully synchronous state $\vp_1=\vp_2=\ldots=\vp_N$, if  the coupling is 
attractive, i.e.
 $|\beta|<\pi/2$, and to remain asynchronous otherwise.

We consider a similar setup for oscillators with an 
internal delayed feedback loop.
The latter is a natural ingredient, e.g, of  lasers with external optical 
feedback~\cite{Masoller-02} 
and of numerous biological systems where signal transmission in 
the feedback pathway may be rather 
slow~\cite{Glass-Mackey-88,*PhysRevLett.85.2026,*Batzel-Kappel-11}. 
It is known, that phase dynamics of an autonomous oscillator with a delayed 
feedback loop 
can be in the simplest case represented as
$\dot\vp=\w+\alpha \sin(\vp_\tau-\vp)$, 
where $\vp_\tau\equiv\vp(t-\tau)$, $\tau$ is the delay, and
$\alpha$ quantifies the feedback 
strength~\cite{Niebur-Schuster-Kammen-91,Masoller-02,%
Goldobin-Rosenblum-Pikovsky-03}.
Assuming the global coupling to be of the Kuramoto-Sakaguchi type
as above, we write our 
basic model as
\begin{equation}
\dot\vp_k=\w+\alpha\sin(\vp_{\tau,k}-\vp_k) +
\e\text{Im}(e^{i\beta}Ze^{-i\vp_k})\;.
\label{eq:ph}
\end{equation}

\begin{figure}[h!]
\centering
\includegraphics[width=0.98\columnwidth]{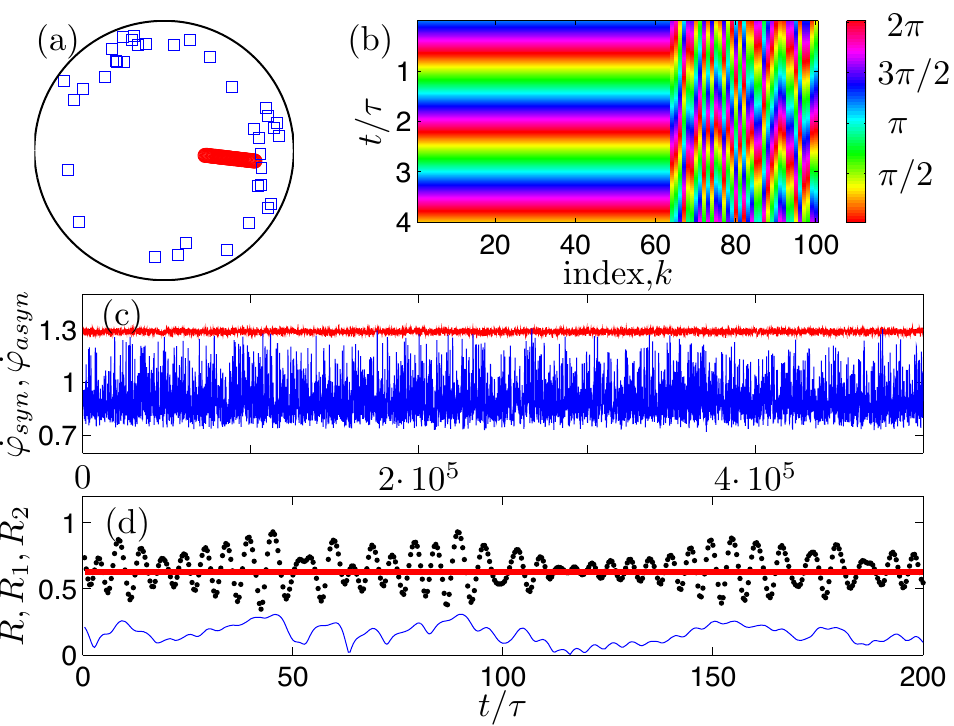}
\caption{Chimera state in model~\eqref{eq:ph}.
(a) Snapshot of the phases reveals that 64 oscillators (red circles, numbered 
with $k=1,\ldots,64$) 
are in the cluster and 36 oscillators (blue squares) belong to the cloud. 
For visibility, the radial coordinate is increased proportionally to the 
oscillator index $k$.
(b) Temporal evolution $\vp_k(t)$, shown by color/grey coding. 
(c) Instantaneous frequencies of an oscillator from the cluster (upper red 
curve) 
and of an oscillator from the cloud (lower blue curve). 
The average values are 
$\langle \dot\vp_{syn}\rangle_t = 1.2897$ (cluster) and 
$\langle \dot\vp_{asyn}\rangle_t = 0.9033$ (cloud).  
(d) Amplitude of the mean field component 
contributed by the cluster, $R_1=|\sum_{k=1}^{64} e^{\vp_k}|/100$ 
(red bold line), and by the cloud, $R_2=|\sum_{k=65}^{100} e^{\vp_k}|/100$ 
 (blue solid line). 
Black dotted line shows the amplitude $R$ of the total mean field.}
\label{fig:chimera}
\end{figure}

We start by numerical demonstration of a chimera-like state 
in model~\eqref{eq:ph} for parameter set $\omega = 1$, $\alpha=1/3$, 
$\beta = \pi/2 + 0.01$, $\tau = \pi-0.02$, $\e = 0.05$, and $N=100$.
In Fig.~\ref{fig:chimera}a,b we show this state after transients in the dynamics 
are over; the snapshot and the time evolution 
of the phases clearly depict a synchronized
cluster of $64$ oscillators and a cloud of $36$ asynchronous ones.  
(Notice that throughout this example we number the oscillators 
in a way that units with 
indices $k=1,\ldots,n$ are in the cluster, whereas units with $k=n+1,\ldots,N$
belong to the cloud.)
Temporal phase dynamics is further illustrated in Fig.~\ref{fig:chimera}c:
for the elements in the cluster it is highly regular with a nearly constant 
instantaneous frequencies,  
while oscillators in the cloud are chaotic and their
instantaneous frequencies strongly fluctuate.
Moreover, individual frequencies
in the cloud are only weakly correlated, so that the phase differences 
demonstrate many phase slips and are unbounded. 
This irregularity is also reflected in the strong 
fluctuations of the cloud contribution to the mean field, to be compared with 
nearly constant contribution from the cluster (Fig.~\ref{fig:chimera}d). 

\begin{figure}[h!]
 \centering
 \includegraphics[width=0.98\columnwidth] {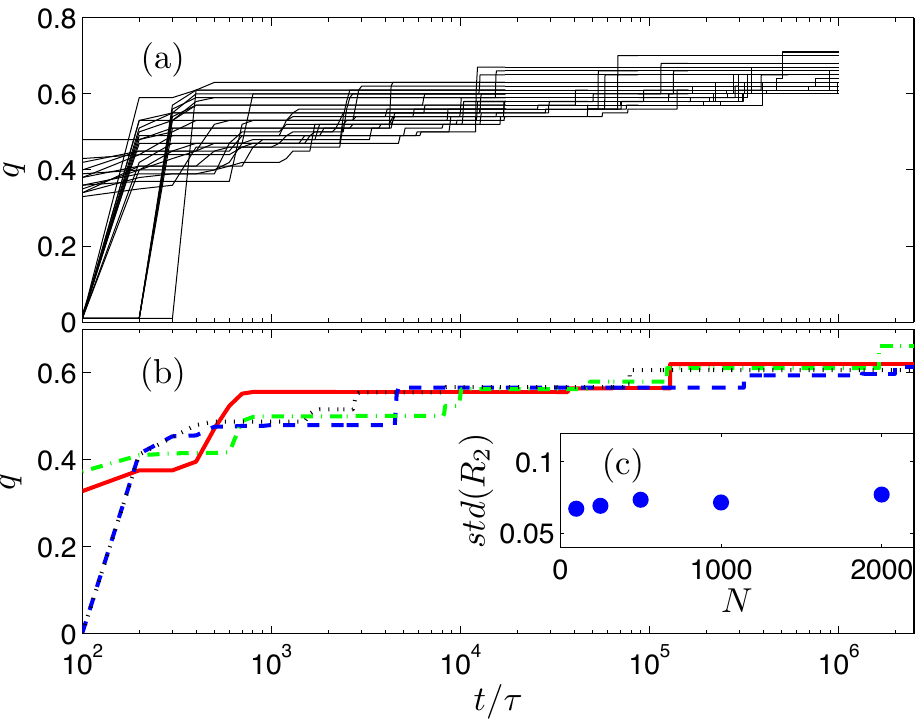}
 \caption{Temporal evolution of the cluster and saturation of its size.
 (a) Growth of the relative cluster size $q=n/N$ for different initial 
 conditions 
 for $N=100$ oscillators.
(b) Saturation of $q$ for different ensemble size: $N=250$ (red solid), $N=500$ (blue dashed),
$N=750$ (green dash-dotted), and $N=1000$ (black dotted). 
(c) Standard deviation for the amplitude of the mean field component $R_2$
contributed by the cloud, for different ensemble size $N$.}
 \label{fig:clgrowth}
\end{figure}

Formation of the chimera state is illustrated in Fig.~\ref{fig:clgrowth}. 
Here in panel (a) we show the 
cluster growth for different initial conditions (different initial cluster size 
and random uniform 
distribution of cloud phases); we see that the cluster size saturates at a 
value between $n=60$ and $n=71$. 
Notice the logarithmic scale of the time axis:
formation of the cluster with $q=n/N\approx 0.5$ is relatively fast, while its further growth 
is an extremely slow process (below we will argue that the full synchrony, i.e. 
the cluster with $q=1$, cannot appear). 

\begin{figure}[h!]
 \centering
 \includegraphics[width=0.95\columnwidth]{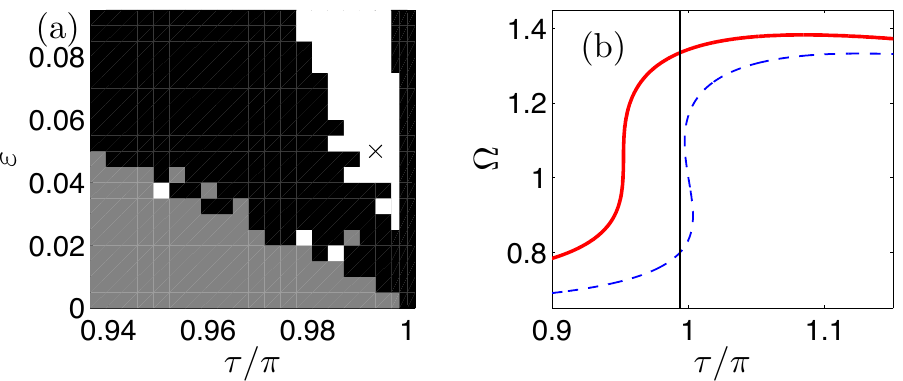}
 \caption{(a) Approximate domain of chimera states (white region); 
 $\w$, $\alpha$, and $\beta$ are 
 same as above, $N=256$. Symbol $\times$ marks the parameters used 
 in Figs.~\ref{fig:chimera},\ref{fig:clgrowth}. In the black domain we observed 
 multi-cluster states, while
 the gray domain corresponds to the states with zero mean field and equal 
 rotation frequencies for all units.
(b) Solution of Eq.~(\ref{freqcluster}): frequency of the one-cluster state 
$\Omega$ as function of $\tau$,
 for uncoupled oscillators, $\e=0$, (blue dashed line)
 and for $\e=0.05$ (red bold line). Vertical black line marks $\tau=\pi-0.02$.
  }
 \label{fig:freq}
\end{figure}

To show that formation of the chimera-like state is not a finite-size effect,  
in Fig.~\ref{fig:clgrowth}b we 
illustrate formation of 
the chimera-like state for ensembles of different sizes, up to $N=1000$.
In all cases the final state has cluster of size $q\approx 0.6$.    
As shown below, for the stability of the chimera-like state it is 
important, that the fluctuation of the order parameter $R_2$ 
of the cloud does not vanish in the thermodynamic limit $N\to\infty$; 
Fig.~\ref{fig:clgrowth}c demonstrates that the variance of $R_2$ practically 
does not depend on $N$ up to values $N=2000$. This fact indicates that 
the units of the cloud are not uncorrelated, but are organized in a collective 
chaotic mode. Finally, we emphasize that chimeras exist not only 
for parameters
chosen above for an illustration, but in a finite 
parameter domain, shown in
Fig.~\ref{fig:freq}a together with domains of other types of dynamics.

Next, we present 
theoretical arguments explaining existence of a chimera-like state  
in model~\eqref{eq:ph}. Let us consider first the fully synchronized,
uniformly rotating one-cluster state $\vp_1=\ldots=\vp_N=\Phi=\Omega t$, 
where frequency 
$\W$ is yet unknown. 
Substituting this expression
into Eq.~(\ref{eq:ph}) we obtain equation 
\begin{equation}
\Omega=\omega-\alpha\sin\Omega\tau+\e\sin\beta\;,
\label{freqcluster}
\end{equation}
its solution  $\Omega(\tau)$  
is shown in Fig.~\ref{fig:freq}b, for cases $\e=0$ (uncoupled oscillators) 
and $\e=0.05$ (one-cluster state). 
We see that in both cases, the solution for the chosen delay $\tau$ is unique, 
i.e. there is no multistability. 
The fully synchronous cluster is, however, unstable. 
Indeed, consider a symmetric 
small perturbation  to two arbitrary oscillators,  $\vp_{1,2}=\Phi\pm\delta$.
Such a perturbation is transversal to the 
synchronization manifold and leaves the mean field 
unchanged; it obeys linearized equation 
$\dot\delta =\alpha\cos(\Omega \tau)(\delta_\tau-\delta)-\e\delta\cos\beta$. 
Most important is the eigenvalue 
which is close to zero; using its smallness we obtain 
in the first approximation 
$\lambda=-\e\cos\beta[1+\tau\alpha\cos(\Omega\tau)]^{-1}$. Because 
for parameters used in Fig.~\ref{fig:chimera}
the quantity in brackets is positive, the fully synchronous 
state for $\e\cos\beta <0$
is unstable.
Physically, this means evaporation of the oscillators from the cluster.
Numerical studies show that the fully 
asynchronous state with uniform distribution of 
phases is unstable, too. 
Although we cannot exclude less trivial asynchronous states, 
i.e. with a non-uniform distribution of phases or with several clusters 
and zero mean field, we have not observed them for the chosen parameters.

A natural question is, 
why a partial cluster with $n<N$ elements (we denote its phase by $\Phi$)
is stable, while 
the 
full synchrony for $n=N$ is not. 
To analyze this, we again denote the perturbed phases of oscillators in the 
cluster
as $\Phi\pm\delta$, and obtain after linearization:
\begin{equation}\label{eq:delta_cluster}
\begin{array}{cl}
 \dot{\delta}(t) = & \alpha \cos(\Phi_\tau-\Phi)(\delta_{\tau}-\delta) - \\
 & - \left[\frac{\e n}{N}\cos\beta + \frac{\e}{N} \sum\limits_{j=n+1}^{N}
  \cos(\vp_{j}-\Phi+\beta)\right] \delta.
\end{array}
\end{equation}
Simultaneously we want to check, whether formation of another cluster via merging of oscillators
from the cloud is possible. For this purpose we assume that two
oscillators in the cloud come close to each other,
so that $\Delta(t)=\vp_k-\vp_l$, $l,k>n$, is small, 
and we can linearize the corresponding 
equations to obtain for the difference
\begin{equation}
\begin{array}{cl}
  &\dot{\Delta}(t) = \alpha\cos(\vp_{l,\tau}-\vp_l)(\Delta_{\tau}-\Delta) - \\
 & - \left[\frac{\e n}{N} \cos(\Phi - \vp_l + \beta) -\frac{\e}{N}\sum\limits_{j=n+1}^{N} 
 \cos(\vp_{j} - \vp_l + \beta)\right] \Delta. \\
\end{array}
\label{eq:Delta}
\end{equation}
We cannot solve Eqs.~(\ref{eq:delta_cluster},\ref{eq:Delta}) analytically, as
$\vp_j(t)$ are unknown irregular functions of time. However,
we solve them numerically for large time interval $T$ together with the full 
system~\eqref{eq:ph} and compute the corresponding Lyapunov exponents 
$\lambda=\lim_{T\to\infty}\frac{\ln\delta(T)}{T}\approx -1.25 \cdot 10^{-2}$
and $\Lambda=\lim_{T\to\infty}\frac{\ln\Delta(T)}{T}\approx 2.38 \cdot 10^{-2}$.  
Because the Lyapunov exponent $\lambda$ describing transversal 
stability of the cluster is negative,
and the exponent $\Lambda$ describing 
transversal stability in the cloud is positive,
 the cluster is stable towards evaporation of the oscillators, while
merging of cloud oscillators to another mini-cluster is forbidden.

Stabilization of the cluster can be qualitatively
explained as follows. 
Contrary to the fully synchronized case, in presence
of a cloud,  oscillators in the cluster are subject to
a force which has two components, as illustrated by Fig.~\ref{fig:chimera}d: 
a regular force from the cluster and an irregular one from the cloud
(last term in Eq.~\eqref{eq:delta_cluster}). 
In the first approximation, the irregular component can be treated as 
a random force, 
and this effective noise is common for all elements of the cluster. 
It is known that common noise tends to 
synchronize oscillators~\cite{Pikovsky-Rosenblum-Kurths-01,Braun-etal-12}. 
Here, for sufficiently strong noise, this 
tendency to synchrony overcomes the internal repulsion 
in the cluster and stabilizes it. However, the cluster cannot 
absorb all elements, because for
$n=N$ the noisy component vanishes; hence, $n<N$. 

Considering now the system from a different 
viewpoint, we discuss, why the periodic forcing 
from the cluster does not entrain the cloud 
oscillators and they eventually do not join the cluster.
Indeed, at initial state of chimera formation more and more oscillators 
join the cluster (see Fig.~\ref{fig:clgrowth}) and the more oscillators 
merge into the cluster, the stronger is the forcing on the cloud oscillators. 
Hence, one may expect the increased tendency to synchrony. However,
with increase of $n$, the frequency of the cluster grows  as described
by  $\Omega=\omega-\alpha\sin\Omega\tau+\e\frac{n}{N}\sin\beta$, 
where in the first approximation we neglect the random forcing from the cloud. 
For $n=64$ the estimated frequency is $\W=1.2901$, 
in a perfect agreement with the observed 
value $1.2897$ (see Fig.~\ref{fig:chimera}c).
Thus, not only the 
amplitude $\e n/N$ of the forcing on non-synchronized units grows with $n$, 
but also the frequency mismatch. 
The growth of the cluster saturates when these values drift outside 
of the synchronization domain for the forced oscillators in the cloud. 
To confirm this, 
we have determined this domain for chosen parameters using a periodic forcing
with parameters taken from the cluster dynamics, and found that 
the forcing with the cluster frequency and the corresponding 
amplitude lies almost exactly at the border of the domain. 
Thus, for $q\approx 65$ further entrainment of oscillators by
the cluster is not possible.

Presented discussion explains 
the mechanism of the \textit{dynamically sustained bistability} 
that underlies the chimera-like state in our globally coupled system
of identical units: the ensembles splits into two parts with completely
different dynamics, and these parts together create a mean field  that
allows such a bistability.
This mechanism is nontrivial, because, as illustrated in Fig.~\ref{fig:freq}b, 
for the chosen parameters the uncoupled
systems are monostable. However, due to interaction, the oscillators become 
effectively 
bistable: being forced by the same field they exhibit two very different 
dynamical patterns. 
The oscillators in one group are regular and therefore easily synchronize
with each other, 
while the others are highly irregular and remain in different asynchronous,
although correlated, states. The global field
that leads to the bistability is dynamically sustained in a self-consistent way.

\begin{figure}[!th]
 \centering
\includegraphics[width=0.88\columnwidth] {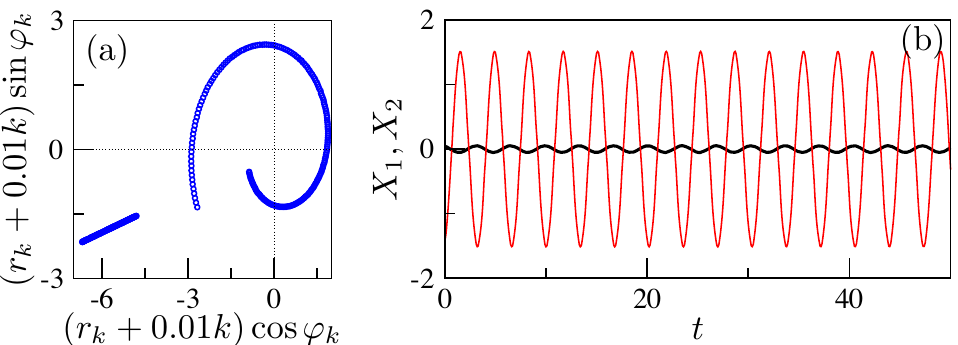}
 \caption{Chimera state in the system of identical Stuart-Landau type oscillators Eq.~(\ref{eq:ls}).
 (a) Snapshot clearly demonstrates one cluster and a group of asynchronous units. 
 Notice that for visibility, in the plot the amplitudes of all units are substituted
as $r_k\to r_k+0.01k$.
 (b) Mean fields of two subgroups, $X_1=N^{-1}\sum_{j=1}^{N/2}r_k\cos\vp_j$ (bold black line)
 and $X_2=N^{-1}\sum_{j=N/2+1}^{N}r_k\cos\vp_j$ (solid red line).
}
 \label{fig:SL}
\end{figure}

Next we discuss less nontrivial, though more transparent, setups where already 
non-coupled oscillators are bistable. Here the coupling is organized in a way, 
that it acts repulsively on the oscillators in one state and attractively on 
those which are in the other state.  
For the first example we consider a model
\begin{equation}
\dot\vp_k=\w+\alpha\sin(\vp_{\tau,k}-\vp_k) +\e R\sin(\Theta_{{\cal T}}-\vp_k+\beta)\;,
\label{eq:ph2d}
\end{equation}
where $R e^{i\Theta} = Z$ and $\Theta({\cal T}) = \Theta(t-{\cal T})$. 
In difference to our model~\eqref{eq:ph}, here not only individual oscillators possess 
a delayed feedback loop, but
the global coupling is also delayed, with another delay time ${\cal T}\ne\tau$. 
Parameters of oscillators are taken as $\w=\pi$, $\tau= 0.99$, and
$\alpha = 1.2$, so that uncoupled units oscillate either with the frequency
$\W_1 = 2.0845$ or $\W_2 = 4.0795$, i.e. are bistable. 
For coupling parameters $\e = 0.1$, $\beta = −\pi/2$, 
and ${\cal T} = 0.2\tau$ we
observe a chimera state (not shown, very similar to the state depicted in Fig.~\ref{fig:SL}), 
what can be explained as follows. 
Suppose there is a non-zero mean field  with the frequency $\nu$.
In the first approximation, the delay in the coupling is equivalent to the
phase shift $\nu {\cal T}$ which sums with the constant 
phase shift parameter $\beta$.
The coupling is attractive if the total shift obeys 
$|\nu {\cal T}+\beta|<\pi/2$, and repulsive otherwise. 
Since the phase shift is frequency-dependent, 
the effective coupling through the same global mean field
is attractive for individual oscillators having frequency $\nu=\W_1$
and repulsive for those with $\nu=\W_{2}$. 
As a result, the sub-population of oscillators which initially are 
in the state 
with $\W_1$ synchronize, while the elements with $\W_2$ remain asynchronous. 

A similar scenario can be implemented with bistable identical 
oscillators without delays. 
Consider  $N$ Stuart-Landau type oscillators, (here written in polar coordinates $r_k,\vp_k$) 
having two stable limit cycles and let these oscillators be globally coupled via an 
additional linear circuit, described by variable $u$:
\begin{equation}
\begin{aligned}
&\dot r_k=0.1r_k(1-r_k^2)(4-r_k^2)(9-r_k^2) + \e \dot u\cos\vp_k\;,\\
&\dot\vp_k=1+\alpha r_k^2 - \e\frac{\dot u}{r_k}\sin\vp_k  \;, \\[-1ex]
&\ddot u +\gamma \dot u +\eta^2 u =N^{-1}\sum_j^N r_j\cos\vp_j \;. 
\end{aligned}
\label{eq:ls}
\end{equation}
Parameters are  $\alpha=0.1$, $\e=0.1$, $\gamma=0.01$, $\eta=1.5$, $N=400$.
In the simulation, initially $N/2$ units were close to the limit cycle with the 
amplitude $\approx 1$ whereas
the others were close to the second limit cycle, with the amplitude $\approx 3$.
The observed chimera state is shown in Fig.~\ref{fig:SL}. 
Indeed, the frequencies of the limit cycle oscillations are $\W_1=1.1$ and $\W_2=1.9$.
Since the resonant frequency of the circuit $\eta$ lies between them,
$\W_1<\eta<\W_2$, the phase shift in the global coupling
introduced by the harmonic circuit is 
attractive for the state with $\W_2$ and repulsive for that with $\W_1$. 

In summary, we have demonstrated numerically and explained 
semi-quantitatively the emergence
of chimera states in ensembles of identical globally coupled oscillators. 
We have outlined a mechanism  of dynamically sustained bistability which results in 
symmetry-breaking of the initially homogeneous system. 
Here, a remarkable constructive role is played by collective chaos
of non-synchronized units: the irregular forcing from the cloud counteracts the 
instability of the fully synchronous state, thus stabilizing the cluster
of synchronized $n<N$ elements.  
We have also demonstrated that chimera-like states are possible
without this mechanism, if the individual units are naturally bistable,
like in setups described by Eqs.~(\ref{eq:ph2d},\ref{eq:ls}).
We stress that the chimera-like regimes here  
are conceptually much simpler than in the model~(\ref{eq:ph}):
the asynchronous oscillators are not chaotic; 
moreover, here the partition
into synchronous and asynchronous states is fully determined by initial conditions,
while in Eq.~(\ref{eq:ph}) the partition appears self-consistently. 
In this Letter we analyzed only ensembles of identical oscillators, as here the effect is mostly 
striking.  However, we expect that the main features survive for
small heterogeneity and/or noise; this issues remain a subject of a future study. 

We acknowledge discussions with K. Krischer, A. Politi, and Yu. Maistrenko. 
A.Y. thanks DFG (grant PI-220/17) for support.
  

%

\end{document}